\documentclass[twocolumn,showpacs,preprintnumbers,prc,superscriptaddress]{revtex4-2}
\usepackage{graphicx,color}
\usepackage{float}
\usepackage{bm}
\usepackage[pdftex,colorlinks=true, linkcolor = blue, citecolor=blue,urlcolor=blue, bookmarksnumbered=true, bookmarksopen=true]{hyperref}
\begin{document}

\title{Consideration of memory of spin and parity in the fissioning compound nucleus by applying the Hauser-Feshbach fission fragment decay model to photonuclear reactions}

\author{T. Kawano}
\email{kawano@lanl.gov}
\affiliation{Los Alamos National Laboratory, Los Alamos, NM 87545, USA}

\author{A.~E. Lovell}
\affiliation{Los Alamos National Laboratory, Los Alamos, NM 87545, USA}

\author{S. Okumura}
\affiliation{NAPC--Nuclear Data Section, International Atomic Energy Agency, Vienna A-1400, Austria}

\author{H. Sasaki}
\affiliation{Los Alamos National Laboratory, Los Alamos, NM 87545, USA}

\author{I. Stetcu}
\affiliation{Los Alamos National Laboratory, Los Alamos, NM 87545, USA}

\author{P. Talou}
\affiliation{Los Alamos National Laboratory, Los Alamos, NM 87545, USA}

\date{\today}
\preprint{LA-UR-22-32277}

\begin{abstract}
Prompt and $\beta$-delayed fission observables, such as the average
number of prompt and delayed neutrons, the independent and cumulative
fission product yields, and the prompt $\gamma$-ray energy spectra for
the photonuclear reactions on $^{235,238}$U and $^{239}$Pu are
calculated with the Hauser-Feshbach Fission Fragment Decay (HF$^3$D)
model and compared with available experimental data. In the analysis
of neutron-induced fission reactions to the case of photo-induced
fission, an excellent reproduction of the delayed neutron yields
supports a traditional assumption that the photo-fission might be
similar to the neutron-induced fission at the same excitation energies
regardless of the spin and parity of the fissioning systems.
\end{abstract}
\maketitle

\section{Introduction}
\label{sec:introduction}
Nuclear fission phenomena have been studied in the three distinct
reaction time domains, namely in the dynamical evolution of compound
nucleus before scission, during the statistical decay of
highly-excited fission fragments, and following the $\beta$-decay of
produced fission products after prompt emission of neutrons and
$\gamma$ rays. The neutrons and $\gamma$ rays emitted from the fission
fragments give particularly important traces to understand this
complicated process as a whole, as the emission of these particles is
governed by the strict spin, parity, and energy conservation rules
during the sequence of fission fragment decay. How the compound
nucleus in a compact configuration splits into two correlated objects,
where the total fission energy is transferred into the excitation
energies of both fragments as well as the kinetic energy of their
relative motion, is one of the common questions in nuclear fission.

Total spin $J$ of the fissioning compound nucleus formed by a
neutron-induced reaction is restricted to ${\bf J}={\bf I} + {\bf j}$,
where $I$ is the target nucleus spin and $j$ is the total spin of the
neutron. For example, a thermal-neutron-induced reaction on $^{235}$U
(ground state spin of $7/2^-$) produces the compound states of $J=3^-$
and $4^-$ only. In contrast to this limited range of $J$, a much wider
distribution of angular momenta in the formed fragments is often
advocated for by both experimental and theoretical
studies~\cite{Wilhelmy1972, Bonneau2007b, Stetcu2013, Bertsch2019,
Wilson2021, Marevic2021, Stetcu2021, Bulgac2021, Bulgac2022} at the
moment of scission. A simple and classical picture of the origin of
the high spin is the torque produced by strong Coulomb repulsion
between the formed fragments, and this repulsion may strongly depend
on the configuration of two compound nuclei in the vicinity of the
scission.  We may deductively infer the configuration of fission
fragments by measuring the characteristics of the emitted neutrons
and/or $\gamma$ rays and comparing them with predictions from recent
fission fragment decay models~\cite{Becker2013, Litaize2010,
Regnier2016, Talou2018, Okumura2018, Talou2021, Lovell2021,
Okumura2022}, where the statistical Hauser-Feshbach
theory~\cite{Hauser1952} ensures the spin-parity conservation at
each stage of the fission fragment decay.

Recently, the Hauser-Feshbach Fission Fragment Decay (HF$^3$D) model
has been successfully applied to neutron-induced fission
reaactions~\cite{Lovell2021, Okumura2022, Kawano2021}, including both
prompt and delayed particle emissions. Although the model depends on
some phenomenological parameters that are often determined by
experimental data such as the initial mass and kinetic energy
distributions of the primary fission fragments, essential properties
of the compound nucleus just after the scission can be studied. The
HF$^3$D model has been applied to neutron-induced reactions, as it
allows us to calculate fission observables in a
system-energy-dependent way. Photo-nuclear fission is an alternative
approach to study fission process~\cite{Silano2018}, where the
electric and magnetic transitions populate selective spin states in
the compound nucleus, different from those populated in $(n,f)$
reactions. By comparing the neutron and photo-induced fission
reactions, we may investigate if information of the spin of the
fissioning compound nucleus is transmitted to the two fission
fragments beyond the sudden transition from one-body compound nucleus.
In first-order approximation, we assume that the photo-nuclear
reaction occurring on the target nucleus $(Z,A)$ at the $\gamma$-ray
energy of $E_\gamma$ is equivalent to the neutron-induced case on
$(Z,A-1)$ at $E_n = E_\gamma - S_n$, where $S_n$ is the neutron
separation energy. We expect that difference in the spin distribution
of the compound system will impact some correlation observables,
although it will be minimal for average quantities like average
multiplicities of prompt particles, and so on.  In this paper, we
provide HF$^3$D analyses for the $\gamma$-ray induced reactions on the
major actinides and discuss the difference and similarity between the
$\gamma$-ray and neutron induced reactions. The HF$^3$D model
calculation is performed with the {\sc BeoH} module, which is part of
the statistical Hauser-Feshbach nuclear reaction code {\sc
CoH$_3$}~\cite{Kawano2019}.  Instead of performing the compound
nucleus decay by applying the Monte Carlo technique~\cite{Becker2013,
Litaize2010, Talou2018, Talou2021}, {\sc BeoH} numerically integrates
the distributions of fission fragment yields, spin, parity, and
excitation energy to produce precise fission observables even when
extremely small fission fragment yields are expected. 
%% The same
%% parameterizations will be ported in the future
%% to \texttt{CGMF}~\cite{Talou2021}, so that correlation observables can
%% be investigated.

\section{Theory}
\label{sec:theory}
\subsection{Spin and parity population in compound nucleus}

The photo-nuclear reaction populates limited space in the spin and
parity states in a compound nucleus. In the giant dipole resonance
(GDR) region (typically around $\sim 10$~MeV for heavy nuclei), the E1
transition is the dominant contribution to the absorption of incident
$\gamma$ rays, which populates $I \pm 1$ states with flipped
parity. The M1 transition, which also populates the $I \pm 1$ states
but with the same parity, is generally weaker in the GDR
region. However, near the neutron separation energy ($\sim 5$~MeV), M1
can be comparable for the deformed systems~\cite{Ullmann2014,
  Guttormsen2014, Mumpower2017}. In contrast, the neutron-induced
reactions in the same energy range lead to wider spin and parity space
in the formed compound nucleus, as the incident neutron can bring more
orbital momenta into the compound.

To illustrate the difference in the populated spin and parity in the
fissioning compound nucleus, we calculate the partial populations for
each $J^\Pi$ to the total compound formation cross section. The partial
compound formation cross section that produces the $J^\Pi$ state is
\begin{equation}
  \sigma(J,\Pi,E) = \frac{\pi}{k^2} g_J \sum_{lj} T_{lj}(E) \ ,
  \label{eq:initpop}
\end{equation}
where $g_J$ is the spin factor, $k$ is the incoming particle
wave number, and $T_{lj}$ is the transmission coefficient for the
orbital angular momentum $l$ and the total spin $j$.  The coupling
scheme is ${\bf J} = {\bf I} + {\bf j}$, and we omit trivial parity
conservation here.  The total compound formation is
\begin{equation}
  \sigma(E) = \sum_{J\Pi} \sigma(J,\Pi,E) \ ,
\end{equation}
and the partial contribution is defined as 
\begin{equation}
  P(E,J,\Pi) = \frac{\sigma(J,\Pi,E)}{\sigma(E)} \ .
\end{equation}

For neutron-induced reactions, the partial population is calculated
from the optical model transmission coefficients. Here we adopt the
coupled-channels optical potential of Soukhovitskii et
al.~\cite{Soukhovitskii2005}. For the $\gamma$-ray incident reactions,
they correspond to the partial contributions from the E1 and M1
transitions. The photo-absorption cross section can be calculated by
the Quasi-particle Random Phase Approximation (QRPA), and we applies
the non-iterative Finite Amplitude Method (FAM)~\cite{Nakatsukasa2007,
  Sasaki2022, Sasaki2022b} based on the Hartree-Fock-BCS
theory~\cite{Bonneau2004} to solve QRPA equations.  From the partial
component of photo-absorption cross section $\sigma_{XL}(E)$, where $X
= {\rm E}$ or M is the type of interaction, and the multipolarity
$L=1$ for the E1 and M1 transitions, the $\gamma$-ray transmission
coefficient $T_{XL}$ is given by
\begin{equation}
T_{XL}(E) = \frac{2}{2l+1}
           \left( \frac{E}{\hbar c} \right)^2 \sigma_{XL}(E) \ .
\end{equation}
By substituting $T_{XL}$ into Eq.~(\ref{eq:initpop}) as $T_{lj} =
T_{LL}$ with a proper parity conservation for the E and M transitions,
$P(J,\Pi)$ is also calculated in the photo-nuclear reaction case.

\begin{figure}
  \begin{center}
    \resizebox{\columnwidth}{!}{\includegraphics{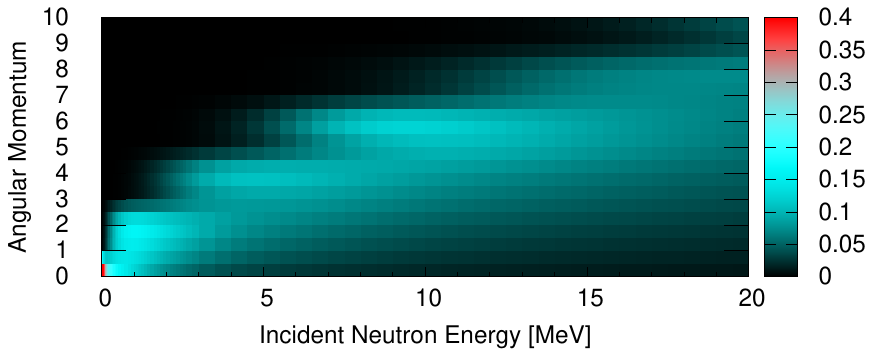}}\\
    \resizebox{\columnwidth}{!}{\includegraphics{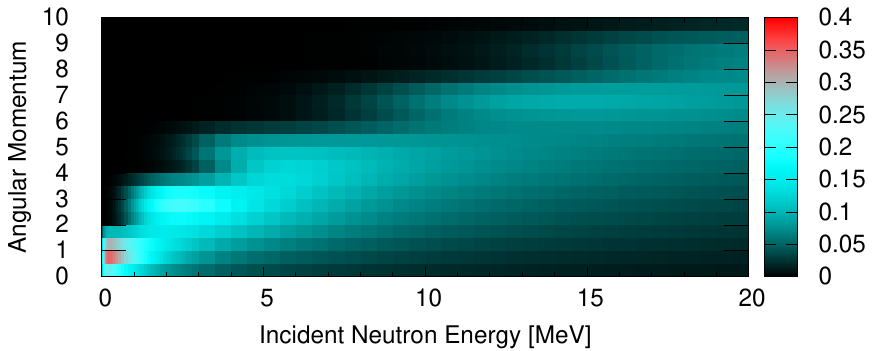}}
  \end{center}
  \caption{Partial population of compound nucleus for the neutron-induced
   reaction on $^{238}$U as a function of neutron energy. The top panel 
   is for the even parity and the bottom is for the odd parity.}
  \label{fig:popinitOM}
\end{figure}

\begin{figure}
  \begin{center}
    \resizebox{\columnwidth}{!}{\includegraphics{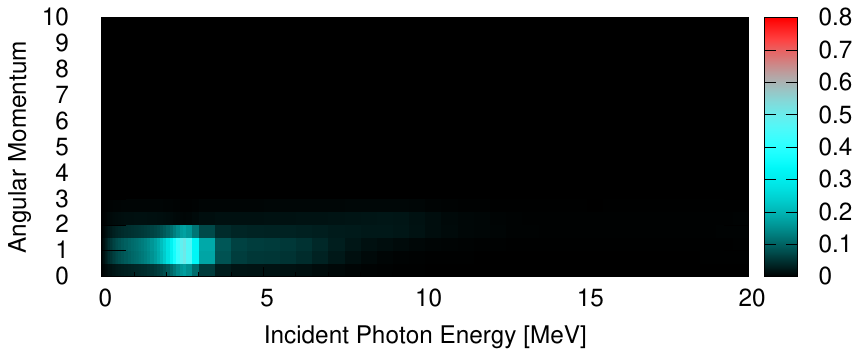}}\\
    \resizebox{\columnwidth}{!}{\includegraphics{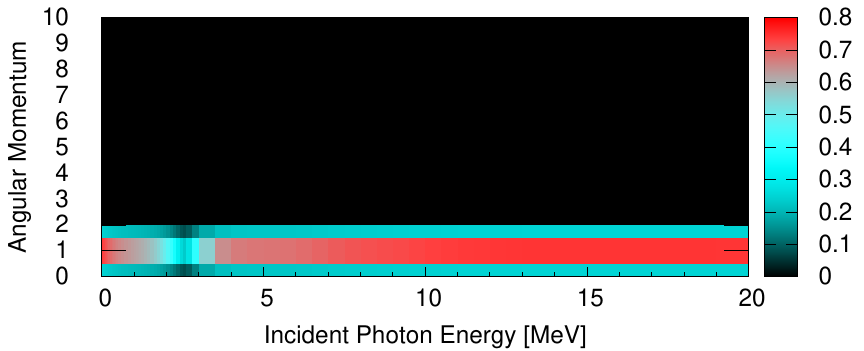}}
  \end{center}
  \caption{The same as Fig.~\ref{fig:popinitOM} but for the photo-induced
   reaction calculated with the QRPA method.}
  \label{fig:popinitGDR}
\end{figure}

Figures~\ref{fig:popinitOM} and \ref{fig:popinitGDR} are the
calculated partial populations with the coupled-channels optical model
and the $\gamma$-ray strength function for the target nucleus of
$^{238}$U. Obviously the compound states formed by a neutron
distribute over a few $\hbar$ width in the GDR region, while $1^{\pm}$
states are selectively populated by the $\gamma$ ray incident.  An
even-odd parity wiggle seen in the population below 5~MeV is due to
competing E1 and M1 transitions, which is also seen in the microscopic
calculations~\cite{Sasaki2022b}.

The population by a $\gamma$ ray can also be estimated by adopting
global parameterization of E1 and M1, such as the generalized
Lorentzian proposed by Kopecky and Uhl~\cite{Kopecky1990} with the
systematic study of M1 scissors mode~\cite{Mumpower2017}.  The result
shows much less parity fluctuation in $P(J,\Pi)$ compared to the QRPA
case.

\subsection{Hauser-Feshbach decay of fission fragments}
\label{subsec:HF3D}

We assume that the two fragments produced by fission are fully
equilibrated compound nuclei after they are well separated from
Coulomb repulsion. These fragments tend to be highly excited and have high
angular momenta. The HF$^3$D model replaces the compound nucleus decay
width by the optical model transmission coefficients in the inverse
reaction channel, as proposed by Hauser and
Feshbach~\cite{Hauser1952}, and the prompt fission observables are
well characterized once the initial configuration of the fission
fragments is specified. Since the details of HF$^3$D model are
discussed elsewhere~\cite{Okumura2018, Lovell2021, Okumura2022}, here
we briefly give an overview of this model. Although we deal with the
photo-induced fission in this study, the formulae are common
unless specified otherwise.

First, we define an initial population $P_{L,H}(E_x,J,\Pi)$ in a
compound nucleus, where the subscript $L,H$ stands for the light and
heavy fragments. The initial population is a joint distribution of
excitation energy $G_{L,H}(E_x)$, spin $R_{L,H}(J)$, and parity
$P(\Pi)$. The population is normalized as
\begin{equation}
\sum_{J\Pi} \int P_{L,H}(E_x,J,\Pi) dE_x = 1 \ .
\end{equation}
We assume the excitation energy distribution is described by Gaussian. The spin
distribution follows the shape given by the Gilbert-Cameron level
density formula~\cite{Gilbert1965, Kawano2013}, but the spin-cutoff
parameter $\sigma^2$ is scaled by an adjustable parameter $f_J$
\begin{equation}
  R_{L,H}(J) = \frac{J+1/2}{(f_J \sigma_{L,H})^2}
              \exp\left\{
                      -\frac{(J+1/2)^2}{2 (f_J\sigma_{L,H})^2}
                  \right\} \ ,
  \label{eq:rjp}
\end{equation}
to take the high spin population in the fission into account. $R(J)$
satisfies the normalization condition of $\sum_{J} R(J) = 1$.
An equal distribution is assumed for the parity distribution.

Second, the total excitation energy $E$ of the system is restricted by
the $Q$-value of splitting the fissioning compound nucleus
$(Z_{\rm CN},A_{\rm CN})$ into the fragments $(Z_L,A_L)$ and
$(Z_H,A_H)$, as well as the total kinetic energy $T$ taken by the
acceleration of fragments. A simple linear dependence on the incident
energy is assumed for the average total kinetic energy
\begin{equation}
  T(E) = T_0 + T_1 E_\gamma \ ,
  \label{eq:TKE}
\end{equation}
where $E_\gamma$ is the incident $\gamma$-ray energy, and the constant
$T_0$ is often taken from a systematic study by
Viola~\cite{Viola1985}, and tuned as necessary. The total excitation
energy is divided into the excitation energy of both the fission
fragments by the so-called $R_T$ model~\cite{Ohsawa1999,Becker2013},
and the distribution width is also determined by the dispersion of $T$
and the kinematics~\cite{Okumura2018}, the initial population
$P_{L,H}(E_x,J,\Pi)$ is thus characterized.

Finally, the aggregation calculation for all the fission fragment
pairs requires distributions of mass and charge numbers. The mass
distribution is modeled by the five Gaussian shapes,
\begin{equation}
 Y_P(A) = \sum_{i=1}^5 \frac{F_i}{\sqrt{2\pi}\sigma_i}
        \exp\left\{
              -\frac{(A-A_m + \Delta_i)^2}{2\sigma_i^2}
            \right\} \ ,
\label{eq:fivegaussian}
\end{equation}
where $\sigma_i$ and $\Delta_i$ are the Gaussian parameters,
$A_m=A_{\rm CN}/2$ is the mid-point of the mass distribution, and
$F_i$ is the fraction of each Gaussian component. We employ Wahl's
$Z_p$ model~\cite{Wahl1988, LA13928} for the charge
distribution. Because Wahl's parameterization of the even-odd term in
the $Z_p$ model is often unsatisfactory to reproduce the delayed
neutron yield~\cite{Minato2018, Okumura2022}, we introduce rescaling
factors of $f_Z$ and $f_N$ to reinforce the even-odd effect
\begin{eqnarray}
  F_Z &=& 1.0 + (F_Z^{\rm W} - 1.0)  f_Z \ , 
  \label{eq:FZ} \\
  F_N &=& 1.0 + (F_N^{\rm W} - 1.0)  f_N \ ,
  \label{eq:FN}
\end{eqnarray}
where $F_Z^{\rm W}$ and $F_N^{\rm W}$ are defined in the original
$Z_p$ model.

\subsection{Model parameters for photo-induced fission}
\label{subsec:parameter}

We have obtained sets of the HF$^3$D parameters for major actinides
($^{235,238}$U and $^{239}$Pu) in our past study~\cite{Okumura2022},
for the neutron-induced fission reactions up to the second chance
fission threshold. The multi-chance fission parameters for $^{235}$U
were also reported by \citet{Lovell2021}. Here we employ the model
parameters obtained by \citet{Okumura2022}, and extend them to the
multi-chance fission calculations for consistency.

Because the mass distribution of Eq.~(\ref{eq:fivegaussian}) is
defined as a relative mass to $A_m$, these distributions can be
applied to each fission chance by shifting $A_m$, {\it e.g.} $A_m =
117.5$ for the first chance of photo-induced fission on $^{235}$U, 117
for the second chance, {\it etc.}  We assume the Gaussian fractions
$F_i$ for the photo-induced fission are the same for the
neutron-induced fission on the same target, but they are corrected by
the neutron separation energy.  We further assume that $R_T$, $f_J$,
$f_Z$, and $f_N$ are the same as the neutron-induced case.  Therefore,
the main difference between the neutron and photo-induced reactions is
the total kinetic energy. Although we are able to adopt the same
values as the neutron-induced data analysis, our preliminary study
showed that the calculated average number of prompt neutrons
$\overline{\nu}_p$ deviates significantly from the experimental data.
Hence, we readjust $T$ to the experimental $\overline{\nu}_p$ in the
next section.

\subsection{Fission observables}

The HF$^3$D model produces several fission observables simultaneously,
such as the average number of prompt neutrons $\overline{\nu}_p$, the
average number of emitted neutrons for each fragment mass
$\overline{\nu}(A)$, the prompt fission neutron and $\gamma$-ray
spectra, and the independent fission product yields $Y_I(Z,A)$, and so
on.  By providing the decay data of fission products, the model also
produces the average number of delayed neutrons $\overline{\nu}_d$ and
the cumulative fission product yields $Y_C(Z,A)$. We take the
evaluated decay data of ENDF/B-VIII.0~\cite{ENDF8} for calculating the
$\beta$-delayed components.

Since the HF$^3$D model has been extended to the multi-chance
fission~\cite{Lovell2021}, in which a few pre-fission neutrons remove
part of total excitation energy, all of the calculated quantities are
incident energy dependent. We calculate these fission observables up
to 20~MeV $\gamma$-ray incident energy, where the typical observed
data are a convolution of $(\gamma,f)$, $(\gamma,nf)$, $(\gamma,2nf)$,
and $(\gamma,3nf)$ reactions.

\section{Results and Discussion}
\label{sec:results}
\subsection{Average number of prompt neutrons}

\begin{figure}
  \begin{center}
    \resizebox{\columnwidth}{!}{\includegraphics{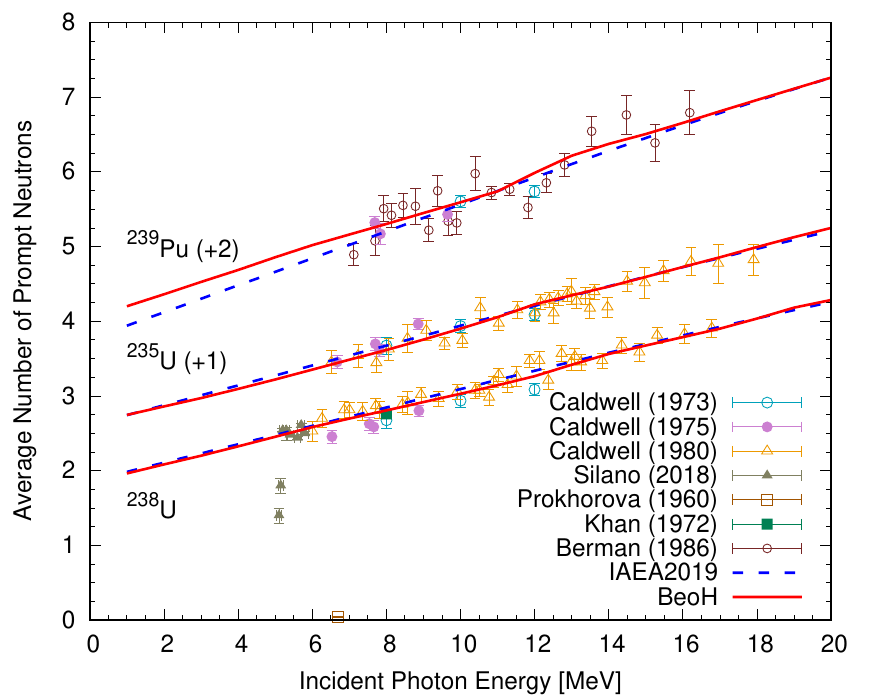}}
  \end{center}
  \caption{The calculated average number of prompt neutrons $\overline{\nu}_p$
   as functions of incident $\gamma$-ray energy compared with the available
   experimental data~\cite{Caldwell1973, Caldwell1975, Caldwell1980, Silano2018, Prokhorova1960, Khan1972, Berman1986}.
   The comparisons for $^{235}$U and $^{239}$Pu are shifted by
   $+1$ and $+2$. The dashed lines are the evaluated $\overline{\nu}_p$ in 
   the IAEA photonuclear data library 2019~\cite{Kawano2020}.}
  \label{fig:nup}
\end{figure}

The average number of prompt neutrons $\overline{\nu}_p$ is sensitive
to the available excitation energies in the light and heavy fragments,
and therefore, it depends on the total kinetic energy,
$T$. Experimental energy dependence of $T$ for major actinides often
shows a negative slope~\cite{ANLNDM64, Zoller1995, Akimov1971,
  Vorobeva1974, Meierbachtol2016} except for a somewhat flattened
behavior observed at very low energies~\cite{Duke2014, Duke2016}. The
linear parameters $T_0$ and $T_1$ in Eq.~(\ref{eq:TKE}) are adjusted
to reproduce the experimental $\overline{\nu}_p$ of the photo-induced
reactions on $^{235,238}$U and $^{239}$Pu. The calculated
$\overline{\nu}_p$'s are compared with experimental
data~\cite{Caldwell1973, Caldwell1975, Caldwell1980, Silano2018, Prokhorova1960,
  Khan1972, Berman1986} in Fig.~\ref{fig:nup}, as well as the
evaluated data in the IAEA photonuclear data library
2019~\cite{Kawano2020}. The IAEA data are identical to the JENDL
photonuclear data file~\cite{JENDL-PD2016} in this energy range.  The
obtained parameters are tabulated in Table~\ref{tbl:Tparm}, and
compared with the systematics of Viola~\cite{Viola1985}. The obtained
parameters for $^{235,238}$U agree fairly well with Viola, while our
$^{239}$Pu is 5.4~MeV higher.  This is due to a smaller $f_J$
parameter of 1.58 for $^{239}$Pu than 2.96 for $^{235,238}$U in
Ref.~\cite{Okumura2022}. The $f_J$ parameter and $T$ are
anti-correlated and different estimates of $f_J$ and $T$ may give
similar results. By comparing to the Viola systematics, a higher $f_J$
might be favorable, nevertheless a definitive conclusion cannot be
cast at this point due to the limited experimental information as well
as the compensation error between $f_J$ and $T$.

In our past analysis of neutron-induced fission~\cite{Okumura2018},
the total kinetic energy for $^{235}$U is estimated to be $T = 171.2 -
0.18E_n$~MeV. By shifting the neutron separation energy of 5.298~MeV
and assuming $T$ weakly depends on the mass, 170.3~MeV is 0.7\% lower
than the obtained $T$ of 171.5~MeV. Although data fitting is not an
objective of this study, such a difference might be compensated by
re-adjusting $f_J$ to obtain the similar quality of fit to
$\overline{\nu}_p$, since $f_J$ changes the competition of neutron and
$\gamma$-ray emission. However, in such a case a survey of the
experimental data on prompt $\gamma$ emission would be in order.

Experimental total kinetic energy data are generally not
mono-energetic. De Clercq et al.~\cite{DeClercq1976} reported
measurements of the total kinetic energy with the bremsstrahlung
$\gamma$-ray source that has the maximum $\gamma$-ray energy of
25~MeV. Their data are $170.6 \pm 2$~MeV for $^{235}$U and $170.9 \pm
2$~MeV for $^{238}$U. The multi-chance fission probability weighted
average of values in Table~\ref{tbl:Tparm} are 169.7~MeV and 171.1~MeV
for $^{235}$U and $^{238}$U, which agree with the experimental data.

The HF$^3$D results slightly deviate from the linear function as
assumed in the IAEA/JENDL evaluations, and this is mainly due to the
multi-chance fission effect. As demonstrated by \citet{Lovell2021},
fission occurs at relatively low excitation energy due to the
multi-chance fission even though the incident particle brings large
energies into the system, and $\overline{\nu}_p$ is determined by the
main fissioning compound nucleus in each multi-chance fission realm.

Note that although we calculate photo-induced reactions for low energy
photons as well, they are often below the fission barriers and these
calculated yields are understood to be for the sub-threshold
fission. In addition to the fact the photo-absorption cross section at
a few MeV is very small, they are hardly observed experimentally.

\begin{table}
  \caption{The estimated total kinetic energies for the photo-fission 
   of $^{235,238}$U and $^{239}$Pu, which are expressed as $T=T_0 + T_1E_\gamma$,
   where $E_\gamma$ is in MeV. They are also compared with the systematics 
   of Viola~\cite{Viola1985}.}
  \label{tbl:Tparm}
  \begin{tabular}{ccrrr}
\hline
Target & CN & $T_0$ [MeV] & $T_1$ & Viola [MeV]\\
\hline
  $^{235}$U &  $^{235}$U  & 171.5  & $-0.10$  & 170.4 \\
           &   $^{234}$U  & 171.6 & $-0.10$  & 170.6 \\
           &   $^{233}$U  & 170.6 & $-0.20$  & 170.8 \\
  $^{238}$U &  $^{238}$U  & 171.1  & $-0.05$  & 169.7 \\
           &   $^{237}$U  & 171.9 & $-0.05$  & 169.9 \\
           &   $^{236}$U  & 172.8 & $-0.05$  & 170.1 \\
  $^{239}$Pu & $^{239}$Pu  & 182.0  & $-0.42$ & 176.6 \\
           &   $^{238}$Pu  & 179.0 & $-0.42$ & 176.8 \\
           &   $^{237}$Pu  & 177.1 & $-0.20$ & 177.1 \\
\hline
  \end{tabular}
\end{table}

\subsection{Prompt $\gamma$-ray energy spectra}

\begin{figure}
  \begin{center}
    \resizebox{\columnwidth}{!}{\includegraphics{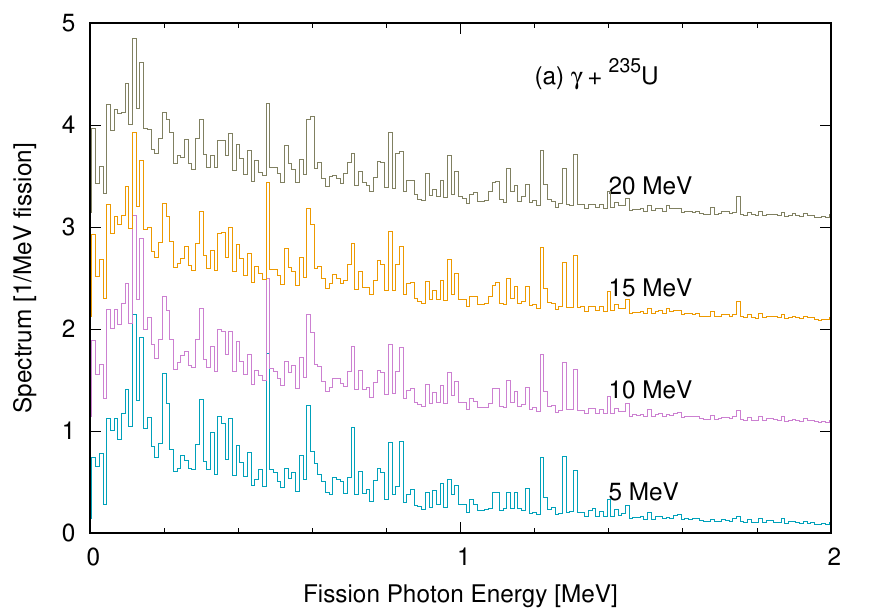}}\\
    \resizebox{\columnwidth}{!}{\includegraphics{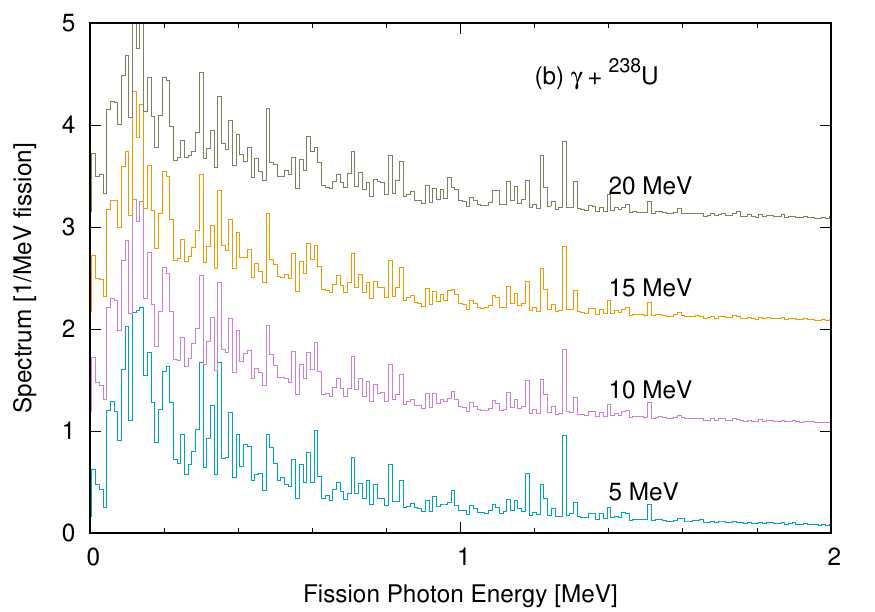}}\\
    \resizebox{\columnwidth}{!}{\includegraphics{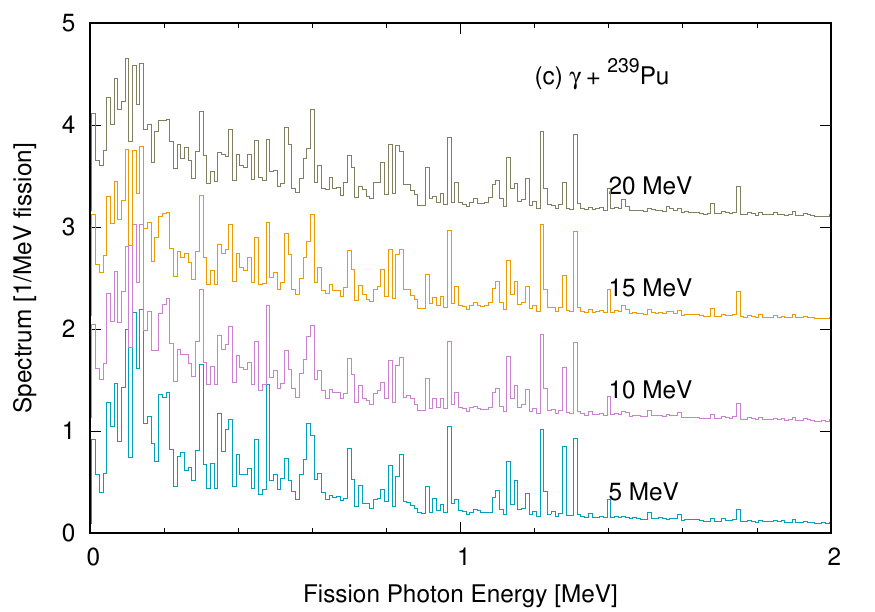}}
  \end{center}
  \caption{The calculated prompt fission $\gamma$-ray spectra for 
   the photo-fission on $^{235,238}$U and $^{239}$Pu at 
   5, 10, 15, and 20~MeV.} 
  \label{fig:gammaline}
\end{figure}

\begin{figure}
  \begin{center}
    \resizebox{\columnwidth}{!}{\includegraphics{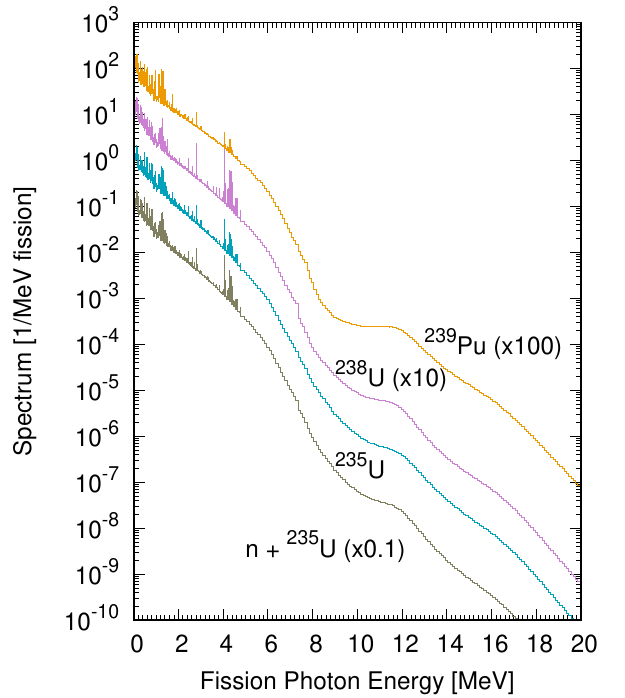}}
  \end{center}
  \caption{The calculated prompt fission $\gamma$-ray spectra for 
   the photo-fission on $^{235,238}$U and $^{239}$Pu at 10~MeV.
   The bottom curve is for the case of thermal-neutron induced fission
   on $^{235}$U. } 
  \label{fig:gammaspec}
\end{figure}

Many prominent discrete prompt fission $\gamma$-ray lines are seen in
the energy spectra, which are produced by the discrete $\gamma$-ray
transitions in the fission products shortly after scission, or
slightly delayed due to the isomer production~\cite{Talou2016,
  Stetcu2014}.  We aggregate the discrete $\gamma$ rays separately
then add them to the continuum spectrum using 10-keV energy bins. This
width roughly corresponds to the Doppler effect caused by the moving
fission fragments. The calculated $\gamma$-ray energy spectra for
selected incident $\gamma$-ray energies are shown in
Fig.~\ref{fig:gammaline} up to 2~MeV. Sometimes a peak in the spectrum
is produced by a single discrete transition, while one histogram bin
may include a few $\gamma$ lines in many cases. The calculated spectra
extend to more than 20~MeV, which is also observed in the
thermal-neutron-induced fission case~\cite{Makii2019}. Their shapes
all resemble one another and are less informative, hence we show the
spectra at the 10~MeV case in Fig.~\ref{fig:gammaspec}. We also
compare these spectra with the case of thermal-neutron-induced fission
on $^{235}$U, where the model parameters are the same as those by
Okumura {\it et al.}~\cite{Okumura2022}.

\subsection{Cumulative fission product yields}

\begin{figure}
  \begin{center}
    \resizebox{\columnwidth}{!}{\includegraphics{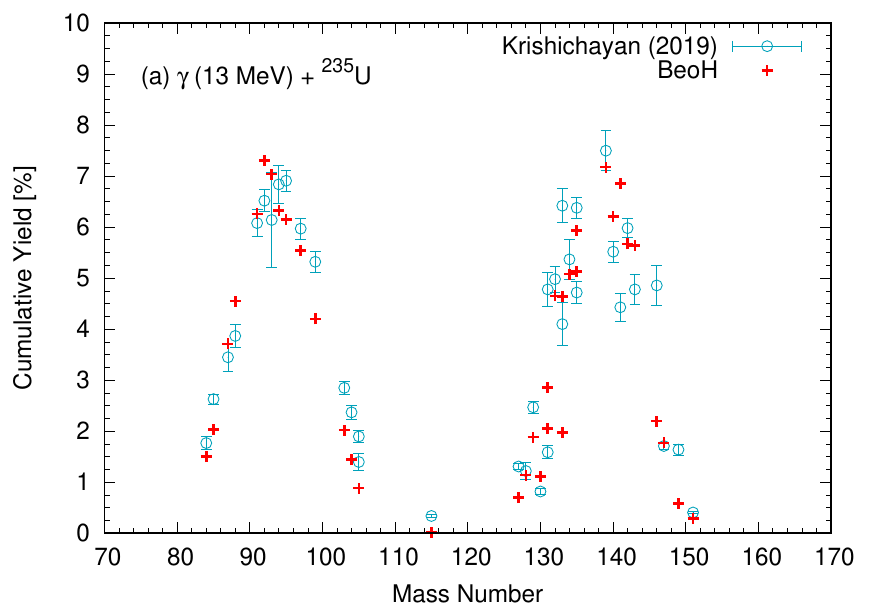}}\\
    \resizebox{\columnwidth}{!}{\includegraphics{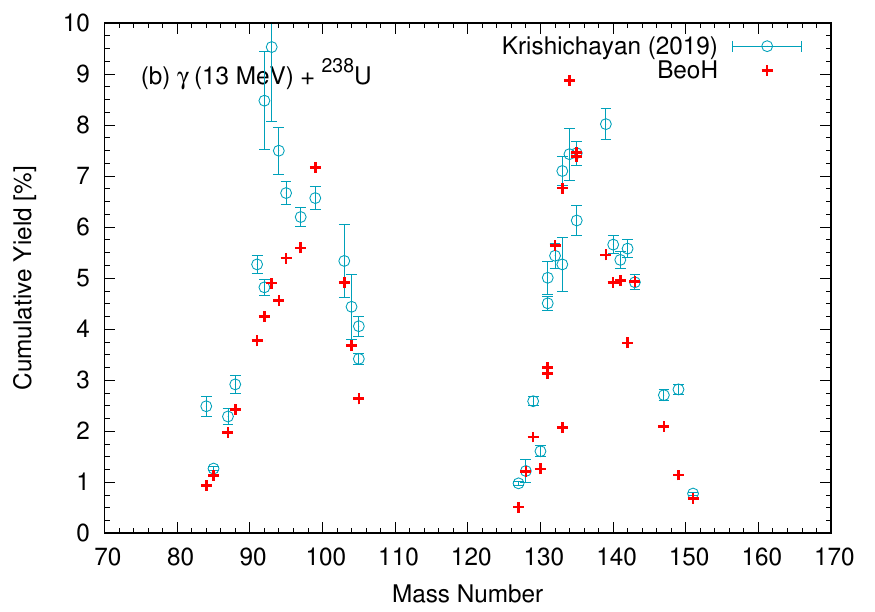}}
  \end{center}
  \caption{Calculated cumulative fission product yields at $E_\gamma = 13$~MeV
           compared with the experimental data of Krishichayan 
           {\it et al.}~\citet{Krishichayan2019}. The top panel (a) is for $^{235}$U,
           and the bottom (b) is $^{238}$U.}
  \label{fig:Krishichayan}
\end{figure}

The experimental data of cumulative fission product yields,
$Y_C(Z,A)$, for photo-fission reactions are rather
inaccessible. \citet{Krishichayan2019} reported 37 $Y_C(Z,A)$ data for
$\gamma + ^{235}$U and 36 for $\gamma + ^{238}$U at 13~MeV those are
measured with the mono-energetic photon beam at the HI$\gamma$S
facility. We compare our calculated results with the data of
\citet{Krishichayan2019} in Fig.~\ref{fig:Krishichayan}.  Here we do
not specify the atomic numbers for the sake of simplicity. Each of the
calculated point corresponds to those measured values. The calculated
$Y_C(Z,A)$ for $^{235}$U agrees with the data reasonably well, while the
case of $^{235}$U has a large deviation in the mass range 90 -- 95. We
cannot reproduce such high $Y_C(Z,A)$ without maintaining rather good
fit to the data in the heavy mass range.

\begin{figure}
  \begin{center}
    \resizebox{\columnwidth}{!}{\includegraphics{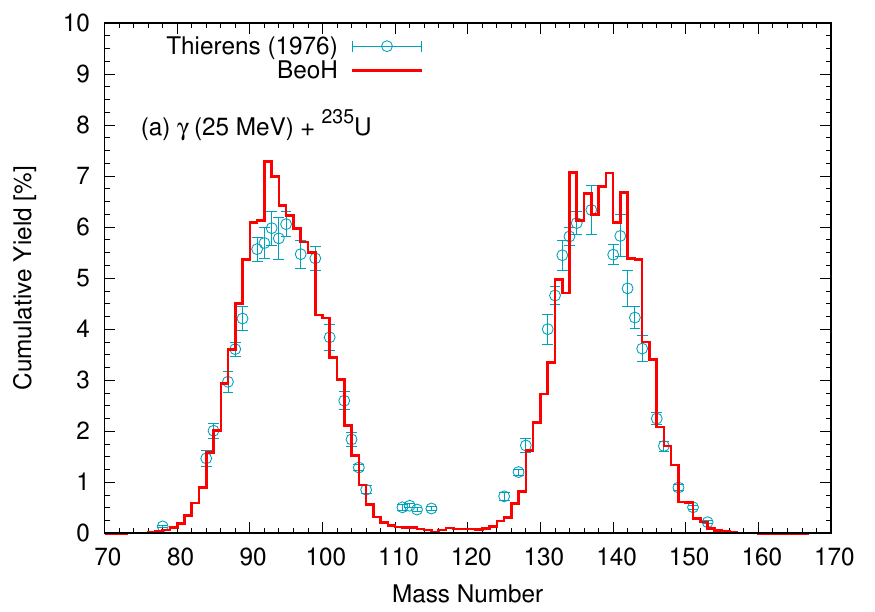}}\\
    \resizebox{\columnwidth}{!}{\includegraphics{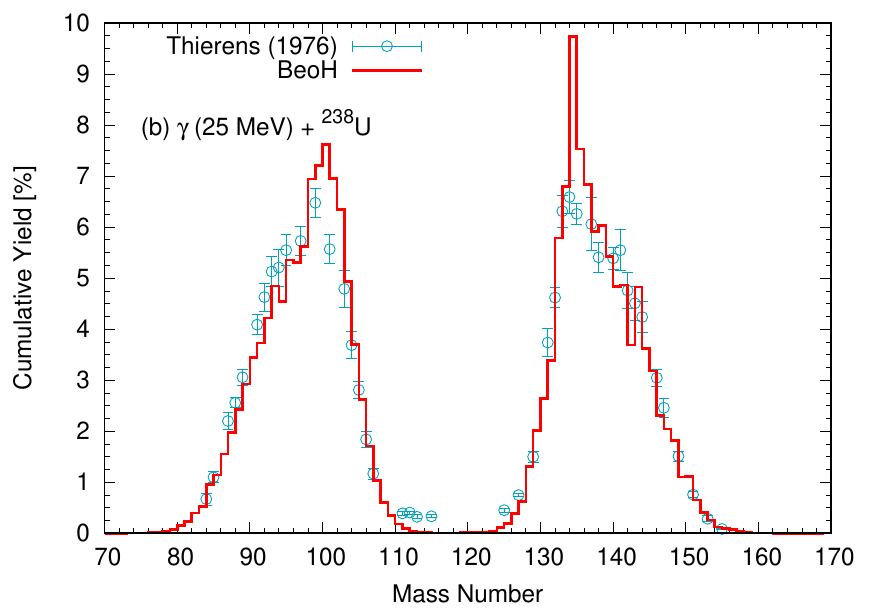}}
  \end{center}
  \caption{Calculated cumulative fission product yields averaged over an estimated
       bremsstrahlung photon energy spectrum that has the maximum energy of
       $E_\gamma = 25$~MeV compared with the bremsstrahlung data of 
       \citet{Thierens1976}. The top panel (a) is for $^{235}$U,
           and the bottom (b) is $^{238}$U.}
  \label{fig:Thierens}
\end{figure}

\citet{Thierens1976} measured the cumulative mass yields by the
bremsstrahlung photon source that has the maximum energy of 25~MeV.
Since an exact photon energy spectrum for this measurement is unknown,
we average our calculated results with the weighting factor of
$w(E_\gamma) = c \sigma_f(E_\gamma) / E_\gamma$, where $\sigma_f$ is
the evaluated photo-fission cross section in the IAEA photonuclear
data library, and $c$ is a normalization factor.  Although this would
not be an exact comparison because of an uncertain photon energy
spectrum at the sample location, Fig.~\ref{fig:Thierens} compares the
energy-averaged mass yield
\begin{equation}
 \overline{Y}_C(A) =
 \int \sum_Z w(E_\gamma) Y_C(Z,A,E_\gamma) dE_\gamma  \ ,
\end{equation}
which seems to be in reasonable agreement except for the symmetric
region, even though the extension of the calculation to 25~MeV is a
rather long-range extrapolation. For better reproduction of the
experimental data, we may need steeper energy dependence of the
symmetric fission component.

\subsection{Average number of delayed neutrons}

The average number of delayed neutrons, $\overline{\nu}_d$, is another
important quantity to validate the calculated $Y_C(Z,A)$.  When a
$\beta$-decay branch of the precursor $(Z,A)$ includes a delayed
neutron emission mode, the delayed neutron yield from this precursor
is calculated as $Y_C(Z,A) b N_d$, where $b$ is the branching ratio to
the neutron-decay mode, and $N_d$ is usually unity unless multiple
neutron emission is allowed. The total delayed neutron yield is thus
$\overline{\nu}_d = \sum Y_C(Z,A) b N_d$, and the summation is
performed on all the delayed neutron emitters. There are about 200 --
300 delayed neutron emitters included in this study.

\begin{figure}
  \begin{center}
    \resizebox{\columnwidth}{!}{\includegraphics{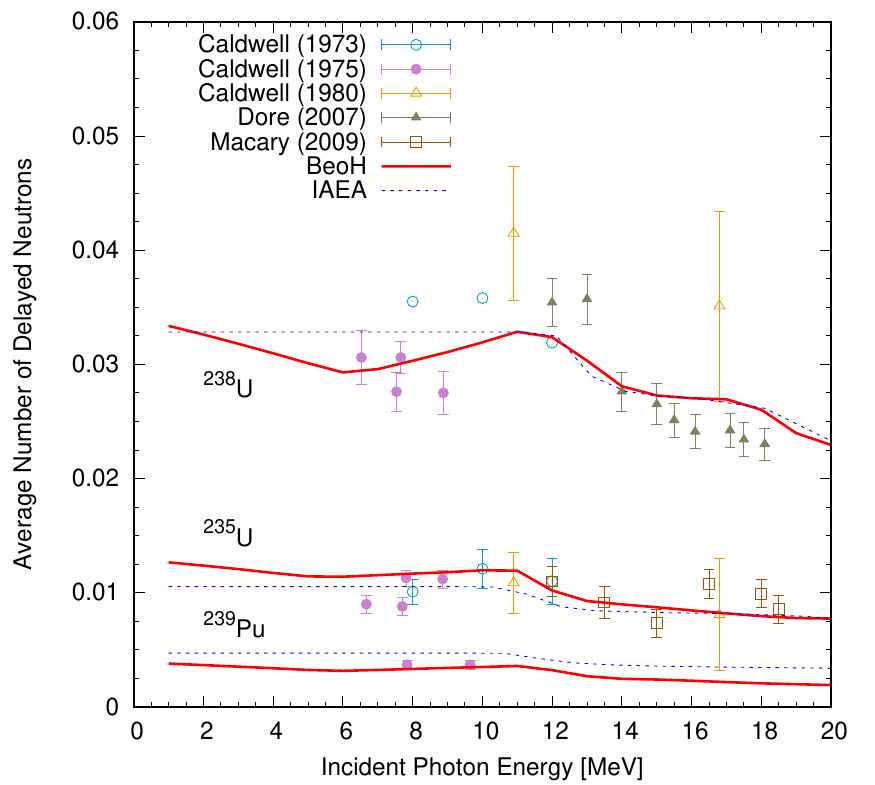}}
  \end{center}
  \caption{The calculated average number of delayed neutrons $\overline{\nu}_d$
   as functions of incident $\gamma$-ray energy compared with the available
   experimental data~\cite{Caldwell1973, Caldwell1975, Caldwell1980,
  Dore2007, Macary2009}. The dashed lines are the evaluated $\overline{\nu_d}$ in 
   the IAEA Photonuclear Data Library 2019~\cite{Kawano2020}.}
  \label{fig:nud}
\end{figure}

The calculated $\overline{\nu}_d$ is shown in Fig.~\ref{fig:nud} as a
function of the incident $\gamma$-ray energy, where the agreement with
the experimental data~\cite{Caldwell1973, Caldwell1975, Caldwell1980,
  Dore2007, Macary2009} is rather remarkable. It should be emphasized
that the calculated $\overline{\nu}_d$ was not fitted to these data,
but they were just consequences of model calculations that are based
on the neutron-induced fission as well as the total kinetic energies
estimated from the average prompt neutrons $\overline{\nu}_p$.  We
also plot the evaluated $\overline{\nu}_d$'s in the IAEA photonuclear
data library in the same figure, which is again the same as JENDL
Photonuclear data file~\cite{JENDL-PD2016}.  It is unclear whether
this evaluation comes from a fit directly to the experimental data or
from a simple functional form. On the contrary, our calculations well
reproduce the data without any further adjustments. Notable structure
in the calculated $\overline{\nu}_d$ is seen near 11 and 17~MeV, which
is clear evidence of the multi-chance fission effects.

As discussed before, the average prompt neutron multiplicity,
$\overline{\nu}_p$, is rather evidence of energy conservation that
determines the available excitation energies for prompt particle
emissions. Even in the multi-chance fission cases, the energy
conservation is still governed by the separation energy of pre-fission
neutrons and the average kinetic energy of the emitted pre-fission
neutrons. Although we modestly adjusted the total kinetic energy $T$
to reproduce the experimental $\overline{\nu}_p$, the changes in $T$
should be within the experimental uncertainties.

On the other hand, many other factors are involved in the delayed
neutron yields, namely the energy dependence of the fission fragment
yields $Y_P(A)$, the even-odd effect, and so on. The multi-chance
fission probabilities change $\overline{\nu}_d$ significantly, because
a shift of one mass unit replaces many of the delayed neutron
precursors. As shown in the decent reproduction of $\overline{\nu}_d$
by applying the neutron-induced fission parameters, it is unlikely
that the produced fission fragments after scission remember the spin
and/or parity of the fissioning system. Our observation supports a
traditional assumption that the photo-fission could be approximated by
the neutron-induced fission at the same excitation energies. This is
in contrast to the very different populations of spin states in the
compound nucleus between the neutron and $\gamma$-ray induced
reactions as shown in Figs.~\ref{fig:popinitOM} and
\ref{fig:popinitGDR}.

\section{Conclusion}
\label{sec:conclusion}
The spin dependence of the nuclear fission was investigated by
applying the statistical Hauser-Feshbach Fission Fragment Decay
(HF$^3$D) model to the photo-nuclear reactions on $^{235,238}$U and
$^{239}$Pu.  We performed the HF$^3$D model calculations for these
targets by employing the model parameters obtained for the
neutron-induced fission cases, with re-adjustment of the total kinetic
energies to reproduce the average number of prompt fission
neutrons. The calculated results were compared with the experimental
cumulative fission product yields at mono-energetic 13~MeV as well as
a continuous bremsstrahlung photon source. We also conducted
comparisons of the average number of delayed neutrons, obtaining 
fairly good agreements with the experimental data. While the populated
spin and parity states in the fissioning system for the neutron and
$\gamma$-ray induced reactions are very different, we found that the
cumulative fission product yields and the average number of delayed
neutrons seem to be insensitive to the spin and parity of the
fissioning systems. This observation supports the traditional
assumption that the photo-fission might be similar to the
neutron-induced fission at the same excitation energies regardless of
the quantum numbers of the fissioning systems.

\section*{Acknowledgments}

TK thanks to Nobuyuki Iwamoto of JAEA for providing the undocumented
information on the JENDL photonuclear data files. He also thanks Anton
Tonchev of LLNL, Matthew Gooden and Gencho Rusev of LANL for valuable
comments on the HI$\gamma$S experimental data, and Nathan Gibson for
encouraging this work. This work was partially support by the Office
of Defense Nuclear Nonproliferation Research \& Development (DNN
R\&D), National Nuclear Security Administration, U.S. Department of
Energy. We gratefully acknowledge partial support by the Advanced
Simulation and Computing (ASC) Program.  This work was carried out
under the auspices of the National Nuclear Security Administration of
the U.S. Department of Energy at Los Alamos National Laboratory under
Contract No. 89233218CNA000001.

%\bibliography{reference}
\bibliography{ref}

\end{document}